\documentclass[11pt]{article}
\usepackage{epsfig}
\usepackage{wrapfig}
\usepackage{subfigure}

\begin{document}

\title{3D Simulation of New Generation Gas Detectors}

\author{S. Mukhopadhyay, N.Majumdar}
\date{\small{INO Section, SINP, Kolkata, India\\
supratik.mukhopadhyay@saha.ac.in, nayana.majumdar@saha.ac.in}}
\maketitle

\begin{abstract}
Simulation of new generation gas detectors is rendered complicated due to
the non-trivial nature of the electric field and simultaneous presence of
several length-scales.
Computation of the electrostatic field, however, is extremely important
since streamers in gas volume and streamers across the dielectric
surfaces are known to cause serious damage to Micro Pattern Gas Detectors (MPGD)
and are the main factors in limiting their gain. In this paper, we present
the use of a nearly exact Boundary Element Method (neBEM) solver that
overcomes most of the limitations of FEM packages and is capable of
producing accurate results.
\end{abstract}

\section{Introduction}
Micro Pattern Gas Detectors (MPGD) were conceived and built as a consequence of
natural evolution from the gas detectors of the earlier generation, namely the
Multiple Wire Proportional Chambers (MWPC) and its many variants. The purpose
was to achieve higher resolution, better stability and higher rate capability
than the earlier detectors. A fairly large number of MPGDs have been developed
and used since the advent of the Micro Strip Gas Chamber (MSGC) in 1988 by Oed
\cite{Oed88}. Importance of detailed detector 
simulation in general, and electrostatics in particular, for understanding the
advantages as well as the disadvantages, becomes quite apparent if we consider
the operating principle of the MPGDs many of which also have complicated
multiple dielectric configuration. In this work, we will deal with MPGDs of
three popular types, namely MSGC, micro MEsh GAseous Structure (microMEGAS)
\cite{Giomataris96}, and Micro Wire Detector (MWD) \cite{Adeva99}.

At present, different steps are
undertaken to carry out detailed simulation of gas detectors.
Among these, we will discuss about the very first crucial
step of computing the electrostatic field which, in the High Energy Physics
(HEP) community, is mostly carried out using commercial package such as Maxwell
\cite{MAXWELL} that uses the finite element method (FEM) to solve for the
electrostatic field for any given geometry and dielectric combination.
In the present work, we present the neBEM solver which uses a completely new
formulation and foundation expressions for implementing the BIE of
electrostatics \cite{EABE2006,NIMA2006}. Through the use of exact analytic
expressions for evaluating the influence of boundary elements,
most of the drawbacks of conventional BEM have
been removed. We hope to show that the solver can be used very effectively to
solve problems related to estimating the electrostatic configuration of
gas detectors, in general and MPGD-s, in paritcular. Towards this end, we
will present line and surface plots of potential and field
and compare them with available results.

\section{Results and discussions}
To demonstrate the advantages of the neBEM solver, we are presenting results
that have been obtained with very coarse discretization, the maximum number
of elements considered being less than 3500.
Here, in order to present the results in the most general
terms possible, we have neither evoked symmetry, nor used any other memory or
computation time saving technique. It may be noted here that, in the following,
our results have been compared with both 2D BEM results (MSGC) and 3D FEM
results obtained using MAXWELL (MWD and microMEGAS).

\subsection{Micro Strip Gas Chamber}
\begin{wrapfigure}[15]{r}{0.5\textwidth}
\begin{center}
\includegraphics[width=0.48\textwidth]{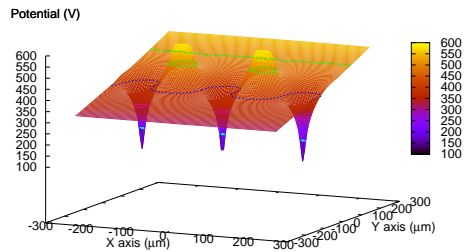}
\caption{\label{fig:MSGCPotSurf} Potential surfaces for a typical MSGC.}
\end{center}
\end{wrapfigure}
The surface plot of electrostatic potential for a typical MSGC
has been presented in Fig.\ref{fig:MSGCPotSurf}.
Qualitatively, the comparison is found to be acceptable with \cite{Randewich94}.
In order to carry out quantitative comparison, we have computed electric field
on the anode for MSGC-s as presented in Table I of \cite{Schmidt94} (cases (a)
 and (b) corresponding to vacuum and
dielectric substrates, respectively). As in \cite{Schmidt94}, the computations
have been done for two thicknesses of the substrate, namely, 100$\mu$m and
500$\mu$m. The electric fields on the anode turn out to be 32.88kV/mm and
35.76kV/mm for a vacuum substrate as the thickness is reduced from 500$\mu$m to
100$\mu$m which according to \cite{Schmidt94} varied from 32.7kV/mm to
36.6kV/mm. Similarly, for a dielectric substrate, the values according to the
present computations are 32.8kV/mm and 37.85kV/mm while those in
\cite{Schmidt94} are 32.5kV/mm and 36.9kV/mm. It may be said that the trend of
the variation is well represented and the numerical values are reasonably close.
The small differences can be attributed to several reasons, the most important
being the fact that the present computations are 3D in nature, while those in
\cite{Schmidt94} were 2D.

\subsection{Micro Wire Detector}
\begin{figure}[hpt]
\centering
\subfigure[Flux contours]{\label{fig:MWDEContour}
	\includegraphics[width=0.45\textwidth]{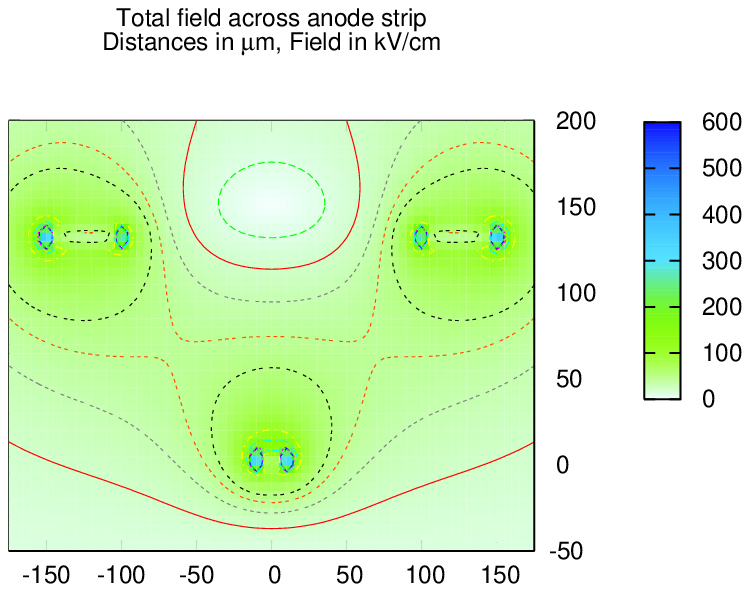}}
\subfigure[Total flux]{\label{fig:MWDTotalE}
	\includegraphics[width=0.45\textwidth]{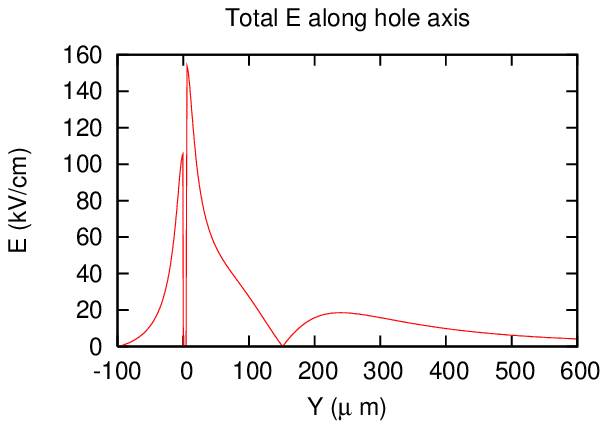}}
\caption{\label{fig:MWDFigures}(a) Flux contour for a microwire detector
and (b) Comparison of total electric field along the hole}
\end{figure}
Here we have considered a typical MWD having the dimensions as in
\cite{Adeva99}. In the following Figs.\ref{fig:MWDEContour} and
\ref{fig:MWDTotalE}, we have presented the contours of the electric field
on the plane perpendicular to the anode axis and the electric field variation
along the hole axis of a typical microwire detector.
Once again, the comparison with \cite{Adeva99} is found to be satisfactory both
qualitatively and quantitatively.

\subsection{microMEGAS}
\begin{figure}[h]
\centering
\subfigure[Flux contours]{\label{fig:ESurfmicroMEGAS}
	\includegraphics[width=0.45\textwidth]{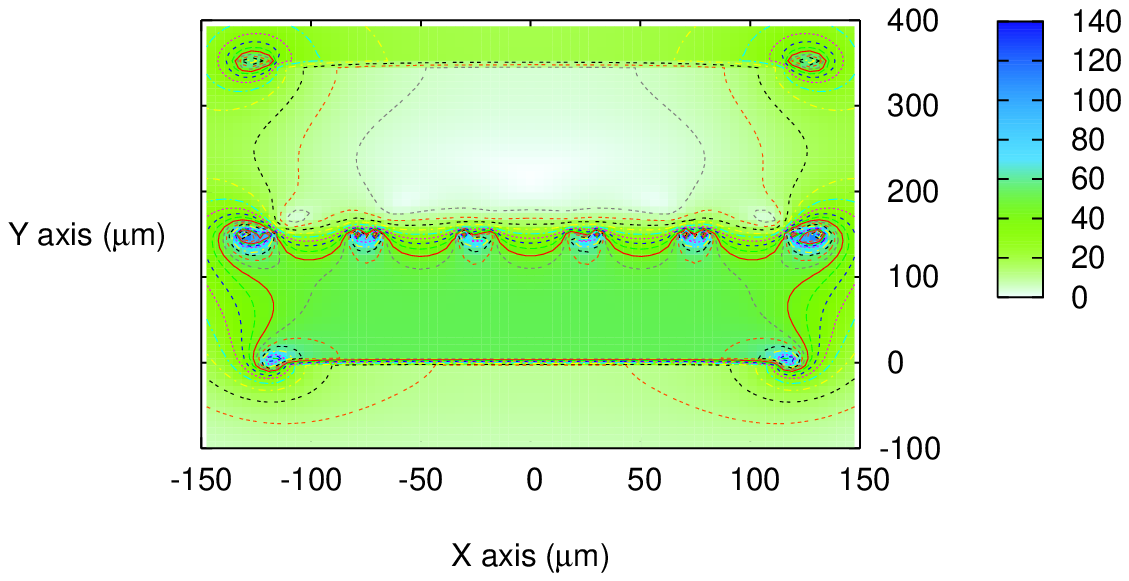}}
\subfigure[Electric field]{\label{fig:ComparemicroMEGAS}
	\includegraphics[width=0.45\textwidth]{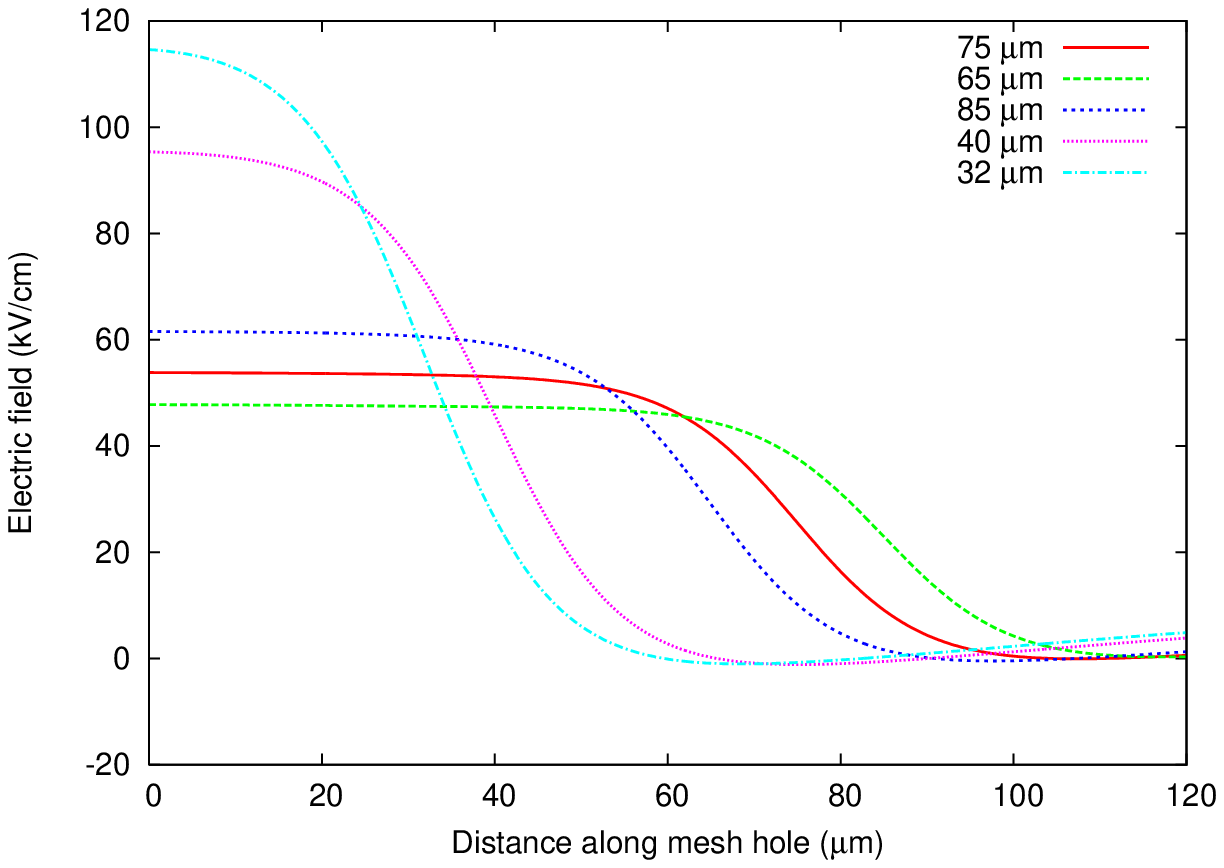}}
\caption{\label{fig:microMEGASFigures}(a) Electric field surface and contours
for a typical microMEGAS (b) Electric field along the axis of a mesh
hole as distance between the mesh and the anode is varied}
\end{figure}
This part of the computation has been carried out in relatively more detail.
We have considered several microMEGAS having geometry as discussed in
\cite{CERNExamples} for easy comparison. In Figs.\ref{fig:ESurfmicroMEGAS} and
\ref{fig:ComparemicroMEGAS}, we have presented the computed electric field
surface and contours and the change in the electric field along the center
of a mesh hole.
These results once again agree with those in \cite{CERNExamples} reasonably
well.

\section{Conclusion}
Using the neBEM solver, it has been possible for us to estimate the
three-dimensional electric field in several micro pattern gas detectors. The
accuracy of the obtained results have been confirmed by comparing them with
existing 2D BEM and 3D FEM results. Despite having a large length scale
variation (1:200) and the use of extremely coarse discretization, the solver
has yielded results that are precise and reliable using little computational
resource. Since detailed simulation of gas detectors begins with the computation
of electrostatic configuration within the device, and depends very critically on
the accuracy of the estimated electric field at any arbitrary point within a
given device, the neBEM solver is expected to become an important tool in
carrying out thorough analysis of gas detectors. This is more true for the
new generation detectors since the length scales of these detectors vary widely
from component to component.

\end{document}